\documentclass[12pt]{spieman}  

\usepackage{amsmath,amsfonts,amssymb}
\usepackage{graphicx}
\usepackage{setspace}
\usepackage[colorlinks=true, allcolors=blue]{hyperref}
\usepackage{tocloft}

\title{New CCD Driving Technique to Suppress Anomalous Charge Intrusion from Outside the Imaging Area for Soft X-ray Imager of Xtend onboard XRISM}

\author[a, *]{Hirofumi Noda}
\author[b]{Mio Aoyagi}
\author[c]{Koji Mori}
\author[d]{Hiroshi Tomida}
\author[e]{Hiroshi Nakajima}
\author[f]{Takaaki Tanaka}
\author[d]{Hiromasa Suzuki}
\author[g]{Hiroshi Murakami}
\author[h]{Hiroyuki Uchida}
\author[h]{Takeshi G. Tsuru}
\author[c]{Keitaro Miyazaki}
\author[c]{Kohei Kusunoki}
\author[d]{Yoshiaki Kanemaru}
\author[i]{Yuma Aoki}
\author[i]{Kumiko Nobukawa}
\author[j]{Masayoshi Nobukawa}
\author[b]{Kohei Shima}
\author[b]{Marina Yoshimoto}
\author[b]{Kazunori Asakura}
\author[b]{Hironori Matsumoto}
\author[k]{Tomokage Yoneyama}
\author[l]{Shogo B. Kobayashi}
\author[m]{Kouichi Hagino}
\author[n]{Hideki Uchiyama}
\author[b]{Kiyoshi Hayashida}
\affil[a]{Astronomical Institute, Tohoku University, 6-3 Aramakiazaaoba, Aoba-ku, Sendai, Miyagi 980-8578, Japan}
\affil[b]{Department of Earth and Space Science, Osaka University, 1-1 Machikaneyama-cho, Toyonaka, Osaka 560-0043, Japan}
\affil[c]{Faculty of Engineering, University of Miyazaki, 1-1 Gakuen Kibanadai Nishi, Miyazaki, Miyazaki 889-2192, Japan}
\affil[d]{Japan Aerospace Exploration Agency, Institute of Space and Astronautical Science, 3-1-1 Yoshino-dai, Chuo-ku, Sagamihara, Kanagawa 252-5210, Japan}
\affil[e]{College of Science and Engineering, Kanto Gakuinn University, Kanazawa-ku, Yokohama, Kanagawa 236-8501, Japan}
\affil[f]{Department of Physics, Konan University, 8-9-1 Okamoto, Higashinada, Kobe, Hyogo 658-8501, Japan}
\affil[g]{Faculty of Informatics, Tohoku Gakuin University, 3-1 Shimizukoji, Wakabayashi-ku, Sendai, Miyagi 984-8588}
\affil[h]{Department of Physics, Kyoto University, Kitashirakawa Oiwake-cho,Sakyo-ku, Kyoto, Kyoto 606-8502, Japan}
\affil[i]{Department of Physics, Kindai University, 3-4-1 Kowakae, Higashi-Osaka, Osaka 577-8502, Japan}
\affil[j]{Faculty of Education, Nara University of Education, Nara, Nara 630-8528, Japan}
\affil[k]{Faculty of Science and Engineering, Chuo University, 1-13-27 Kasuga, Bunkyo, Tokyo 112-8551, Japan}
\affil[l]{Department of Physics, Faculty of Science, Tokyo University of Science, Kagurazaka, Shinjuku-ku, Tokyo 162-0815, Japan}
\affil[m]{Department of Physics, University of Tokyo, 7-3-1 Hongo, Bunkyo, Tokyo 113-0033, Japan}
\affil[n]{Science Education, Faculty of Education, Shizuoka University, Suruga-ku, Shizuoka, Shizuoka 422-8529, Japan}

\begin{document} 
\maketitle

\begin{abstract}

The Soft X-ray Imager (SXI) is an X-ray CCD camera of the Xtend system onboard the X-Ray Imaging and Spectroscopy Mission (XRISM), which was successfully launched on September 7, 2023 (JST).
During ground cooling tests of the CCDs in 2020/2021, using the flight-model detector housing, electronic boards, and a mechanical cooler, we encountered an unexpected issue. Anomalous charges appeared outside the imaging area of the CCDs and intruded into the imaging area, causing pulse heights to stick to the maximum value over a wide region.
Although this issue has not occurred in subsequent tests or in orbit so far, it could seriously affect the imaging and spectroscopic performance of the SXI if it were to happen in the future.
Through experiments with non-flight-model detector components, we successfully reproduced the issue and identified that the anomalous charges intrude via the potential structure created by the charge injection electrode at the top of the imaging area.
To prevent anomalous charge intrusion and maintain imaging and spectroscopic performance that satisfies the requirements, even if this issue occurs in orbit, we developed a new CCD driving technique. This technique is different from the normal operation in terms of potential structure and its changes during imaging and charge injection.
In this paper, we report an overview of the anomalous charge issue, the related potential structures, the development of the new CCD driving technique to prevent the issue, the imaging and spectroscopic performance of the new technique, and the results of experiments to investigate the cause of anomalous charges.

\end{abstract}


{\noindent \footnotesize\textbf{*}Hirofumi Noda,  \linkable{hirofumi.noda@astr.tohoku.ac.jp} }

\vspace{0.1cm}
\section{INTRODUCTION}
\label{sec:intro}  

The X-Ray Imaging and Spectroscopy Mission (XRISM) is the 7th X-ray astronomical satellite of Japan, successfully launched on September 7, 2023 (JST). It is equipped with two soft X-ray observational instruments\cite{Tashiro2024}.
One is the soft X-ray spectroscopic telescope, Resolve, which combines an X-ray Mirror Assembly (XMA)\cite{Hayashi2024} and an X-ray microcalorimeter\cite{Kelley2024}. The other is the soft X-ray imaging telescope, Xtend, which combines an XMA\cite{Tamura2024} and an X-ray CCD camera\cite{Mori2024}.
The X-ray microcalorimeter of Resolve achieves unprecedentedly high energy resolution compared to previous X-ray detectors, allowing it to resolve fine spectral features that are blurred in conventional X-ray observations.
Xtend achieves a wide field of view (FOV) of $38' \times 38'$ in the 0.4--13~keV range, which is crucial for maximizing the spectroscopic performance of Resolve. For example, Xtend can precisely evaluate the contamination in the Resolve FOV ($2.9' \times 2.9'$) from outside sources by utilizing its wide FOV.
When Resolve observes a point source within a diffuse object, emission from the diffuse object might contaminate the spectrum. Xtend can assess this contamination, allowing accurate extraction of target signals from a precise spectrum of Resolve.
In addition, Xtend can complete some orbital operations and calibrations that Resolve alone cannot, such as the initial alignment of the XMA central axis and detector aim point. The pixel size of Resolve is $832~\mu$m, whereas the CCD camera of Xtend has a pixel size of $24~\mu$m, enabling Xtend to achieve more precise alignment than the PSF core of XMA.
Naturally, Xtend is powerful not only in combination with Resolve but also on its own, capable of observing broadly extended diffuse sources and discovering new transient objects in its wide FOV, even during a single pointing observation.

\begin{figure} [t]
   \begin{center}
   \begin{tabular}{c} 
   \includegraphics[height=9cm]{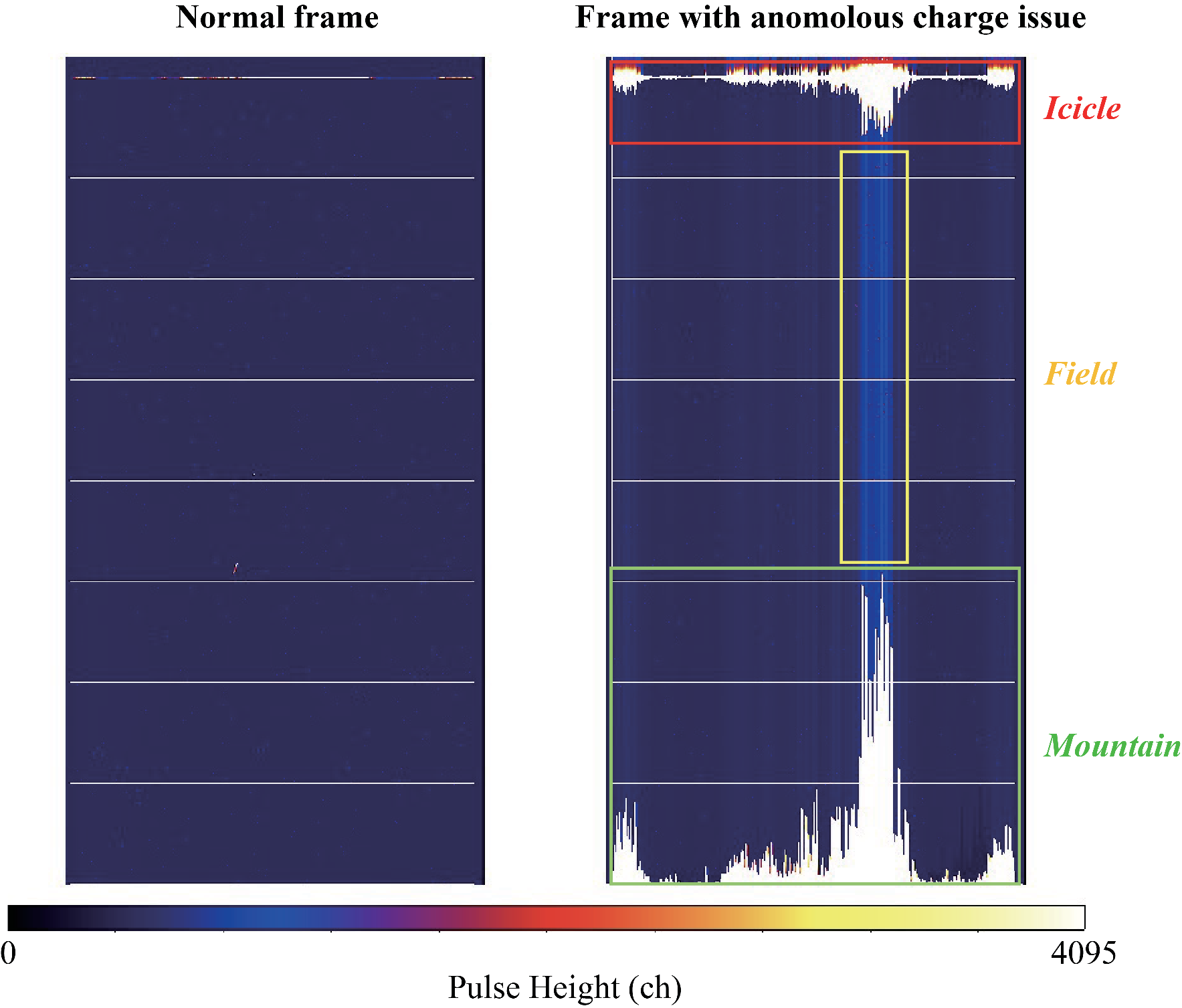}
   \end{tabular}
   \end{center}
   \caption[example] 
   { \label{fig:issue} 
Comparison of a normal raw frame image and a frame image affected by the anomalous charge issue, both without X-ray irradiation. These were derived by the same segment of the same CCD chip at $-110~^{\circ}$C. The color scale represents the pulse height values. We designate the red, yellow, and green square regions as ``\textit{Icicle},'' ``\textit{Field},'' and ``\textit{Mountain},'' respectively. 
}
   \end{figure}

We have been developing the X-ray CCD camera for Xtend, named the ``Soft X-ray Imager (SXI)'',  with a $2 \times 2$ CCD array. 
As shown in Hayashida et al. (2018)\cite{Hayashida2018} and Nakajima et al. (2020)\cite{Nakajima2020}, we included several improvements to increase the optical blocking performance and charge transfer efficiency in the design of the CCD (PchNeXT4) chips of the Hitomi/SXI, based on their performance in orbit\cite{Nakajima2018}. 
From 2017 to 2019, we performed design-verification experiments using small-size prototypes of the improved chips, named PchNeXT4A. Note that the signal carieer is a hole in both PchNeXT4 and PchNeXT4A chips.
Subsequently, we conducted screening experiments to select four CCD chips from twelve Flight-Model (FM) candidates provided by Hamamatsu Photonics K.K. We performed on-ground calibration experiments using the CCD driving system with the multi-color X-ray generator at Osaka University\cite{Yoneyama2020} from 2019 to 2020.
In 2020/2021, we combined the selected four CCDs with the FM detector housing, electronic boards, and a mechanical cooler, and performed an on-ground cooling test of the SXI sub-system alone. During this test, an unexpected issue occurred.
When the CCD chips were cooled and raw frame images were taken without X-ray irradiation, 
the resulting frame images showed unusual features that are not seen in normal frame images (Figure~\ref{fig:issue}).
Note that this issue was not confirmed in Hitomi/SXI on the ground\cite{Tanaka2018} or in orbit\cite{Nakajima2018}.
We then selected four FM chips from the remaining FM candidates and replaced the chips that experienced the issue with these new ones. After the replacement, this issue has not occurred with the FM chips in the SXI on-ground tests\cite{Mori2022} and initial operations in orbit\cite{Suzuki2024}.
The CCD chips installed after the replacement have the same design as those that experienced the anomalous charge issue, suggesting that the issue is unlikely to be specific to these chips.
 
Although the anomalous charge issue has not occurred to date with the FM chips, if it were to happen in the future, the imaging and spectroscopic performance of the SXI would be significantly affected. Therefore, we need countermeasures to prevent this issue and conduct investigations into its causes.
For this purpose, we reproduced the issue in an experimental room using the chips that experienced the issue and a non-FM system, and developed a new CCD driving technique to prevent it. Furthermore, we investigated the source of anomalous charges outside the imaging area of a CCD chip through cooling experiments with the non-FM system.
In this paper, we explain the potential shapes created by the electrodes for the Charge Injection (CI) technique, which are closely related to this issue, in \S\ref{sec2}. Then, we provide an overview of the anomalous charge issue in \S\ref{sec3}. In \S\ref{sec4}, we summarize the development of the new CCD driving technique to prevent the issue and the imaging and spectroscopic performance achieved with the new technique. Finally, in \S\ref{sec5}, we describe the experiments to investigate the cause of anomalous charges outside the imaging area.
The results of the present paper were previously published as an SPIE proceedings paper\cite{Noda2024}.

\section{Electrodes for Charge Injection Technique}
\label{sec2}  

   \begin{figure} [t]
   \begin{center}
   \begin{tabular}{c} 
   \includegraphics[height=8cm]{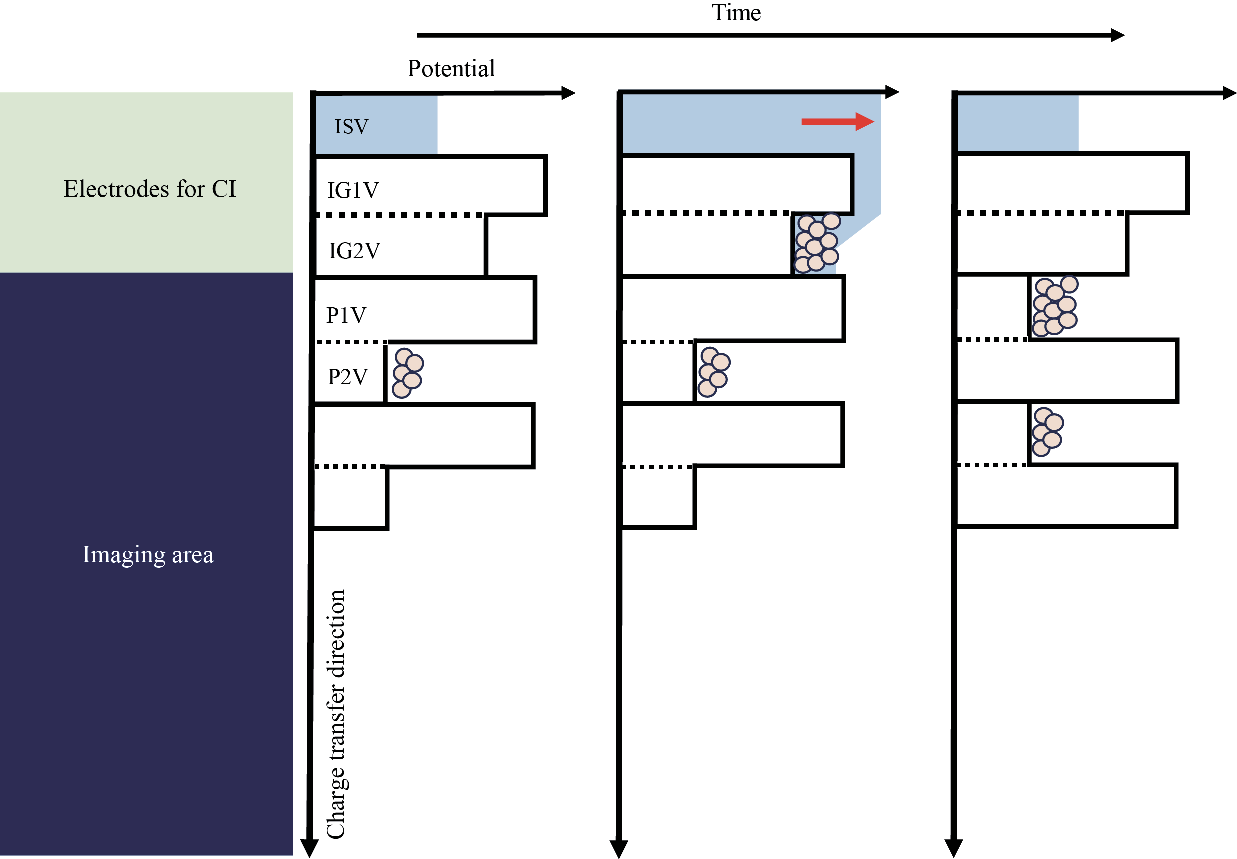}
   \end{tabular}
   \end{center}
   \caption[example] 
   { \label{fig:ci} 
Schematic picture illustrating the potential shapes and their variations for the CI technique used in SXI. }
   \end{figure} 

SXI employs the frame-transfer architecture, and the PchNeXT4A chips have both the Imaging Area (IA) and the Frame Storing (FS) region, each with $1280 \times 1280$ physical pixels. By $2\times2$ on-chip binning, an obtained image has $640 \times 640$ logical pixels. Each chip is equipped with four readout nodes, and two of them are used to read out signals from the entire IA. 
Half of the IA is referred to as a ``segment,'' and each segment uses a single readout node.
Normally, pedestal pulse height values in a raw frame image are tuned to $300 - 700$~channels (ch), except for the CI lines saturated at 4095~ch, which occur every 80 logical pixel (160 physical pixel) lines as shown in Figure~\ref{fig:ci} (left).  
The CI technique is utilized to maintain better spectroscopic performance in orbit. If lattice defects appear due to cosmic rays, etc., the charge being transferred may be captured, leading to a decrease in the number of charges read out relative to the energy of the incident X-ray\cite{Prigozhin2008, Nakajima2008}. This results in a broadening of the spectrum and a deterioration of spectroscopic performance. Therefore, charge traps are filled in advance by artificially injecting charges to prevent the capture of the transferring charge.
The interval of 80 logical pixels between two adjacent CI lines was optimized to maintain spectroscopic performance while minimizing the reduction of the imaging area.

The CI function of SXI is carried out by three electrodes named Input Source Voltage (ISV), Input Gate-1 Voltage (IG1V), and Input Gate-2 Voltage (IG2V), which are attached to the top of the CCD device on the opposite side of the charge transfer direction. A conceptual diagram illustrating the potential of each electrode during CI is shown in Figure~\ref{fig:ci}.
The gate electrodes IG1V and IG2V have structures where the potential of IG1V is higher than that of IG2V. Then, for every 80 logical pixel rows transferred, the potential of the drain electrode ISV is raised once, and artificial charges are injected into the IA via the potential structure created by IG1V and IG2V. The amount of CI is adjusted to ensure that the peak pulse height value reaches the saturation level with normal gain. Through this technique, a sufficient amount of charges is injected into all columns of the IA. The quantity of injected charges is adjusted to be less than that determined by a certain fraction of the full-well, thereby preventing charge leakage in the vertical transfer direction.

\section{Anomalous charge issue}
\label{sec3}  

During the SXI cooling test in 2020/2021, the anomalous charge issue occurred, and raw frame images obtained at $-110$~$^{\circ}$C were markedly different from normal ones, even though no X-rays were irradiated, as shown in Figure~\ref{fig:issue} (right). 
To varying degrees, anomalous charges occurred in all four CCD chips in the array, and they did not disappear throughout the cooling test. While the structures of the regions with high pulse heights varied between different CCD chips, they were nearly identical across all frame images obtained from the same CCD chip.

They exhibited two prominent areas where pulse heights were saturated at the maximum value of 4095~ch, indicated by the red and green square regions in Figure~\ref{fig:issue} (right). Additionally, streaks with pulse heights higher than the pedestal values were also observed, as indicated by the yellow square region in Figure~\ref{fig:issue} (right). Hereafter, we refer to the top (red) and bottom (green) regions saturated at 4095~ch as ``\textit{Icicle}'' and ``\textit{Mountain},'' respectively, and the streaks in the middle field (yellow) as ``\textit{Field}'' for simplicity.

   \begin{figure} [t]
   \begin{center}
   \begin{tabular}{c} 
   \includegraphics[height=7.5cm]{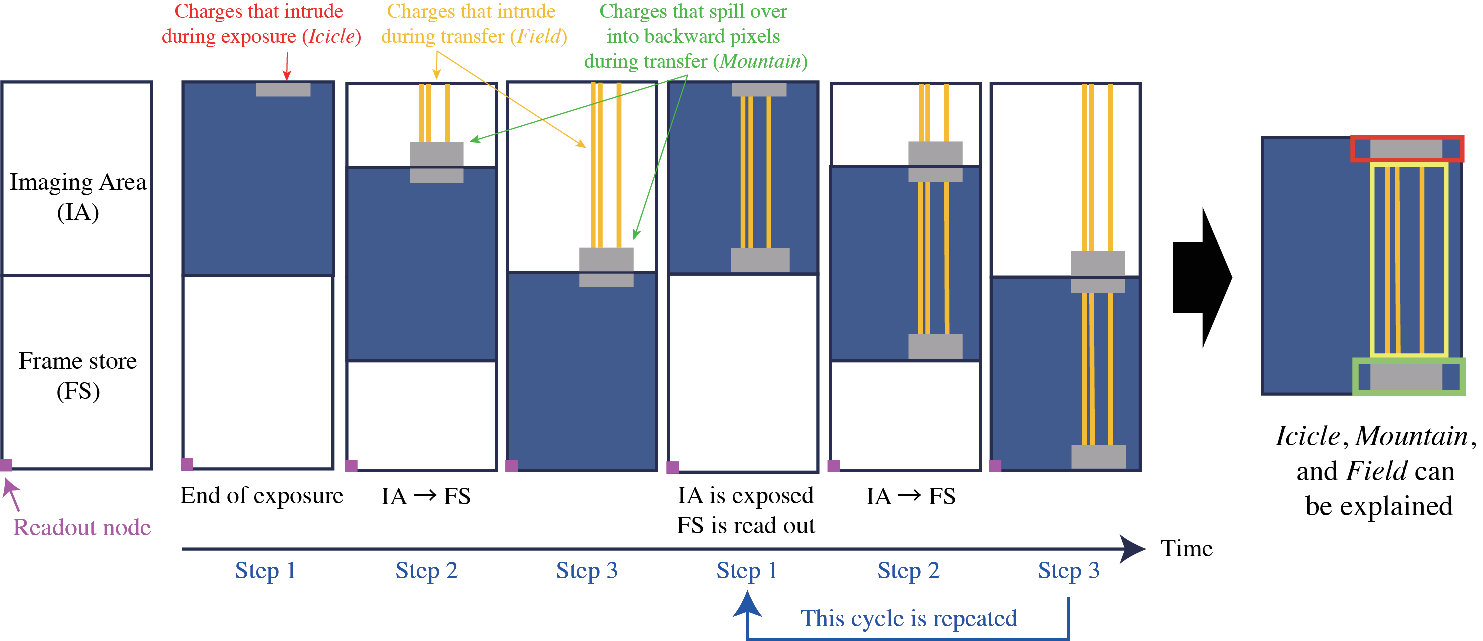}
   \end{tabular}
   \end{center}
   \caption[example] 
   { \label{fig:mechanism} 
A model explaining all the \textit{Icicle}, \textit{Field}, and \textit{Mountain} structures observed in a raw frame image by charge intrusion from the top of a CCD chip. The time flow of signal readout in a segment of a CCD chip is shown, with purple squares indicating the position of the readout node.}
   \end{figure} 

While we have not identified the cause of anomalous charges, we have successfully comprehended all the \textit{Icicle},  \textit{Field}, and \textit{Mountain} structures using a model that combines continuous charge intrusion from outside to the top of CCD chips and vertical charge transfer from IA to FS, as following steps. 

\begin{enumerate}
\item[(Step~1)] Anomalous charges intruding from outside IA exceed the charge storage capacity (full-well) and overflow into multiple pixels. The degree of charge intrusion varies among columns. This results in the appearance of \textit{Icicle}.
\item[(Step~2)] During the vertical charge transfer from IA to FS, a large fraction of the anomalous charges that intruded during exposure remain untransferred in backward pixels because the amount of charges that can be transferred from IA to FS is approximately 20\% of the full-well by design. In addition, anomalous charges continue to intrude during transfer, which appear as \textit{Field}.
\item[(Step~3)] The anomalous charges that remain untransferred in backward pixels are not read out in this frame cycle but in the next frame cycle, which appear as \textit{Mountain} in the next frame. Because the 80\% of the charges that cannot be transferred forms the \textit{Mountain}, the height ratio between the \textit{Icicle} and the \textit{Mountain} becomes approximately 1:4.
\end{enumerate}
These steps are illustrated in Figure~\ref{fig:mechanism}. 
Here, the charge transfer time per physical pixel in the vertical direction is $14.4~\mu$seconds, while the transfer time per logical pixel in the horizontal direction is $28.8~\mu$seconds.
One of the two readout nodes for the segment was used to obtain the frame images, as shown in Figure~\ref{fig:mechanism}. Switching to the other readout node did not eliminate the anomalous charges, which were observed at comparable levels.
As seen above, the model assuming that the anomalous charges intrude solely through the top of a CCD well explains all the characteristic features in frame images. This motivated us to develop a new CCD driving technique to prevent charge intrusion by combining the potentials of IG1V, IG2V, and P1V, which are located at the top of CCD chips.

\section{New CCD driving technique against anomalous charge issue}
\label{sec4}

   \begin{figure} [t]
   \begin{center}
   \begin{tabular}{c} 
   \includegraphics[height=14cm]{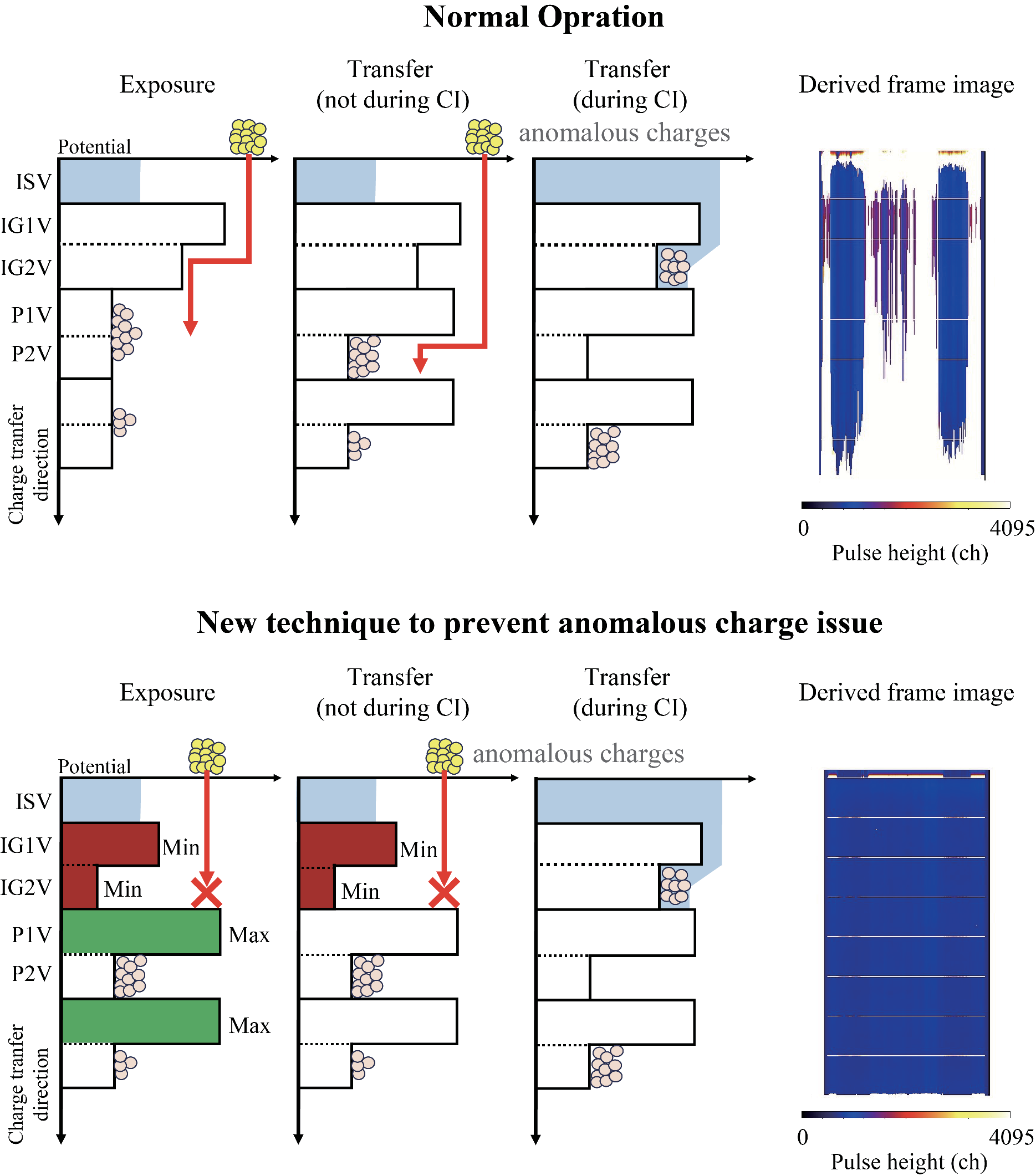}
   \end{tabular}
   \end{center}
   \caption[example] 
   { \label{fig:new} 
(Top) A conceptual diagram illustrating potential shape changes and a resulting frame image when the CCD temperature is approximately $-45~^{\circ}$C, derived in the cooling experiment by the non-FM system at Osaka University with the normal CCD operation of SXI. (Bottom) Same as the top panel but with the new CCD driving technique to prevent the anomalous charge issue. 
}
   \end{figure} 

   \begin{figure} [t]
   \begin{center}
   \begin{tabular}{c} 
   \includegraphics[height=5cm]{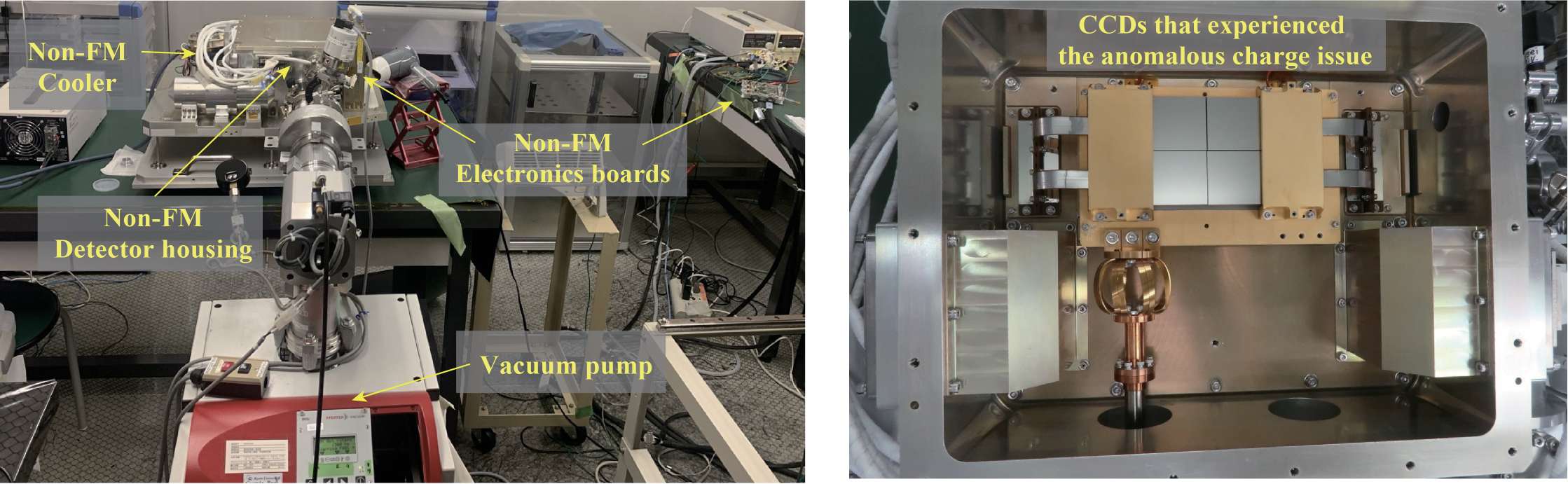}
   \end{tabular}
   \end{center}
   \caption[example] 
   { \label{fig:setup} 
(Left) A picture of the experimental setup to reproduce the anomalous charge issue at Osaka University. (Right) A picture of the interior of the detector housing shown in the left panel. }
   \end{figure} 

With normal CCD operation, the potential of P1V is minimized and kept lower than those of IG1V and IG2V during exposure and at certain times during the clocking for vertical transfer, as shown in Figure~\ref{fig:new} (top left). Consequently, anomalous charges from outside IA cannot be blocked by the potentials of IG1V and IG2V, and they can intrude into IA from the top of CCD chips.
To prevent charge intrusion, one simple method is to minimize the potentials of IG1V and IG2V and maximize that of P1V, creating a large potential barrier against anomalous charges from outside of IA. However, in the ordinary operation of SXI, the potentials of IG1V and IG2V do not change, and they will always remain smaller than the potential of P1V.
With such a potential structure, artificial charges cannot be injected into IA every 80 logical pixel lines, and the CI technique cannot be applied. Therefore, if the potentials of IG1V and IG2V are simply kept minimized, the spectroscopic performance of SXI will be significantly worse than that of normal operation.

To achieve both prevention of charge intrusion and high spectroscopic performance, we developed a new CCD driving technique as follows:
\begin{itemize}
\item During exposure, the potential of P1V is maximized, while those of IG1V and IG2V are minimized, creating a large potential barrier against anomalous charges from outside of IA.
\item When charges are vertically transferred in IA and FS, the potentials of IG1V and IG2V are kept at a minimum except during the CI timings. Meanwhile, the potential of P1V swings for vertical charge transfer.
\item Only during the CI timings every 80 logical pixel lines, the potentials of IG1V and IG2V return to their normal states to inject artificial charges.
\end{itemize}
With this method, the spectroscopic performance is maintained by the application of the CI technique, while the intrusion of anomalous charges is prevented by the large potential barrier, as shown in Figure~\ref{fig:new} (bottom left). Although anomalous charges can intrude into IA when the potentials of IG1V and IG2V return to nominal values in the CI timings, each potential returning lasts only $\sim 5$~milliseconds, and the amount of charge intrusion is considered to be small enough.

   \begin{figure} [t]
   \begin{center}
   \begin{tabular}{c} 
   \includegraphics[height=7.5cm]{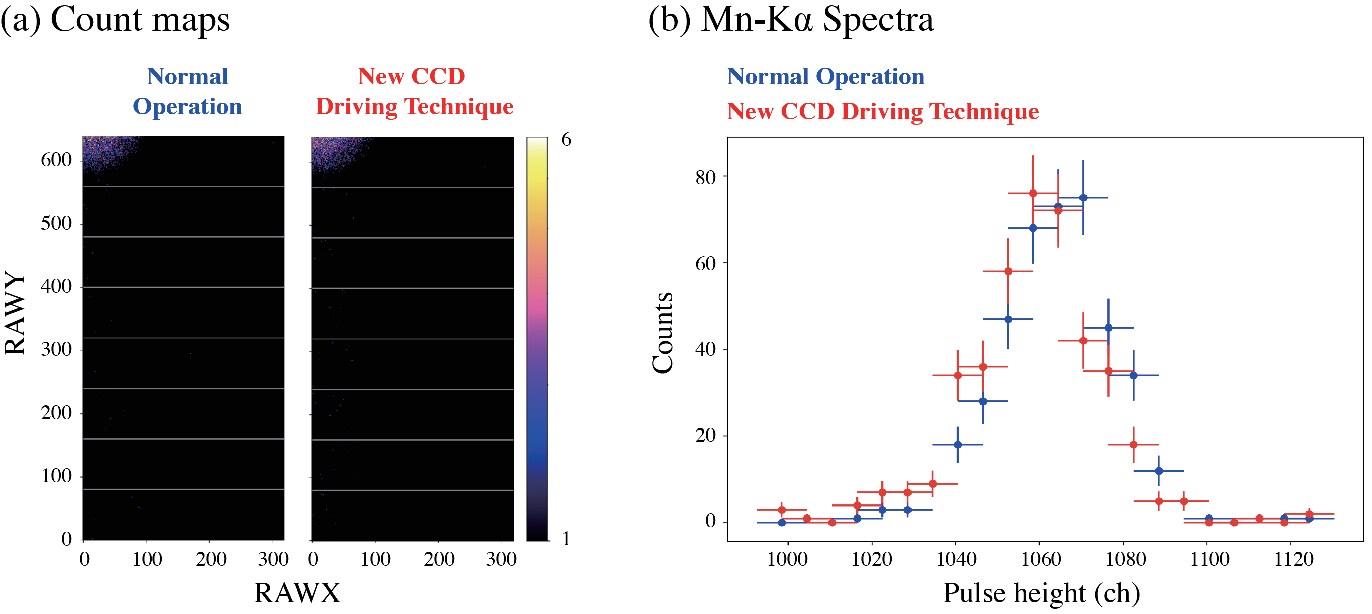}
   \end{tabular}
   \end{center}
   \caption[example] 
   { \label{fig:tvt} 
(a) Count maps obtained by the normal operation and the new CCD driving technique during the satellite thermal vacuum test in 2022. (b) Mn-K$\alpha$ spectra derived in the satellite thermal vacuum test. Blue and red shows those obtained with the normal operation and the new CCD driving technique, respectively. The count maps and spectra were derived using the same segment of the same CCD chip.  }
   \end{figure} 

To test the new CCD driving technique, we constructed an experimental setup by combining the CCD chips that experienced the anomalous charge issue during the SXI cooling test 2020/2021 with a non-flight-model (non-FM) detector housing, electronics boards, and mechanical cooler, as shown in Figure~\ref{fig:setup}.
Using this non-FM system, we conducted CCD cooling tests and successfully reproduced the anomalous charge issue with normal CCD driving operation. 
It should be noted that the anomalous charge issue was not reproduced every time we cooled the CCDs with this non-FM system. When anomalous charges did occur, they were observed in all cooled CCDs, albeit to varying degrees. The trigger for this issue remains unclear.
The anomalous charge issue exhibits a temperature dependence, worsening at higher temperatures (see \S\ref{sec5p1}). 
In this experiment, the CCD temperature was maintained at approximately $-45~^\circ$C to ensure that even more severe anomalous charge issue than that observed in Figure~\ref{fig:issue}~(right) could be effectively suppressed.
Figure~\ref{fig:new} (top right) displays a frame image affected by the reproduced anomalous charge issue. 
The reproduced \textit{Icicle} and \textit{Mountain} features were more severe than those observed during the SXI cooling test in 2020/2021 (Figure~\ref{fig:issue} right), with \textit{Icicle} connected to \textit{Mountain}, and pulse heights in many columns stuck at 4095~ch from top to bottom, overlapping \textit{Field}. 
We applied our new CCD driving technique and confirmed that almost all of the \textit{Icicle}, \textit{Mountain}, and  \textit{Field} structures disappeared, as shown in Figure~\ref{fig:new} (bottom right). Therefore, we concluded that the potential barrier created by the difference between IG1V/IG2V and P1V is sufficient to prevent the intrusion of a large number of anomalous charges from outside IA. We also confirmed that the duration of returning the potentials of IG1V and IG2V to their normal values for the CI technique is short enough to suppress the intrusion of anomalous charges.

We also tested the new CCD driving technique on the FM CCD chips with the FM detector housing, electronics boards, and mechanical cooler during the satellite thermal vacuum test in 2022. In this test, the spacecraft was placed in a thermal vacuum chamber, and its thermal environment was changed like in orbit\cite{Mori2022}. We cooled the FM CCD chips to $-110$~$^{\circ}$C and confirmed that the anomalous charge issue did not occur. Although we could not test the prevention of charge intrusion under these conditions, we could verify the imaging and spectroscopic performance of the new CCD driving technique by irradiating 5.9~keV Mn-K$\alpha$ photons onto the CCD chips from the $^{55}$Fe calibration sources.
Figure~\ref{fig:tvt} displays the count map and Mn-K$\alpha$ spectrum obtained by one segment with normal CCD driving operation in blue. The imaging and spectroscopic performance at 5.9~keV were confirmed to meet the requirements. Subsequently, we applied the new CCD driving technique and obtained a count map and Mn-K$\alpha$ spectrum consistent with those from normal operation, as shown in red in Figure~\ref{fig:tvt}. Therefore, the new CCD driving technique successfully functioned on the FM system, and its imaging and spectroscopic performance were confirmed to be comparable to that of normal operation.

\section{Investigation into the Cause of Anomalous Charges}
\label{sec5}

\subsection{Dependence on CCD Temperature}
\label{sec5p1}

   \begin{figure} [t]
   \begin{center}
   \begin{tabular}{c} 
   \includegraphics[height=7cm]{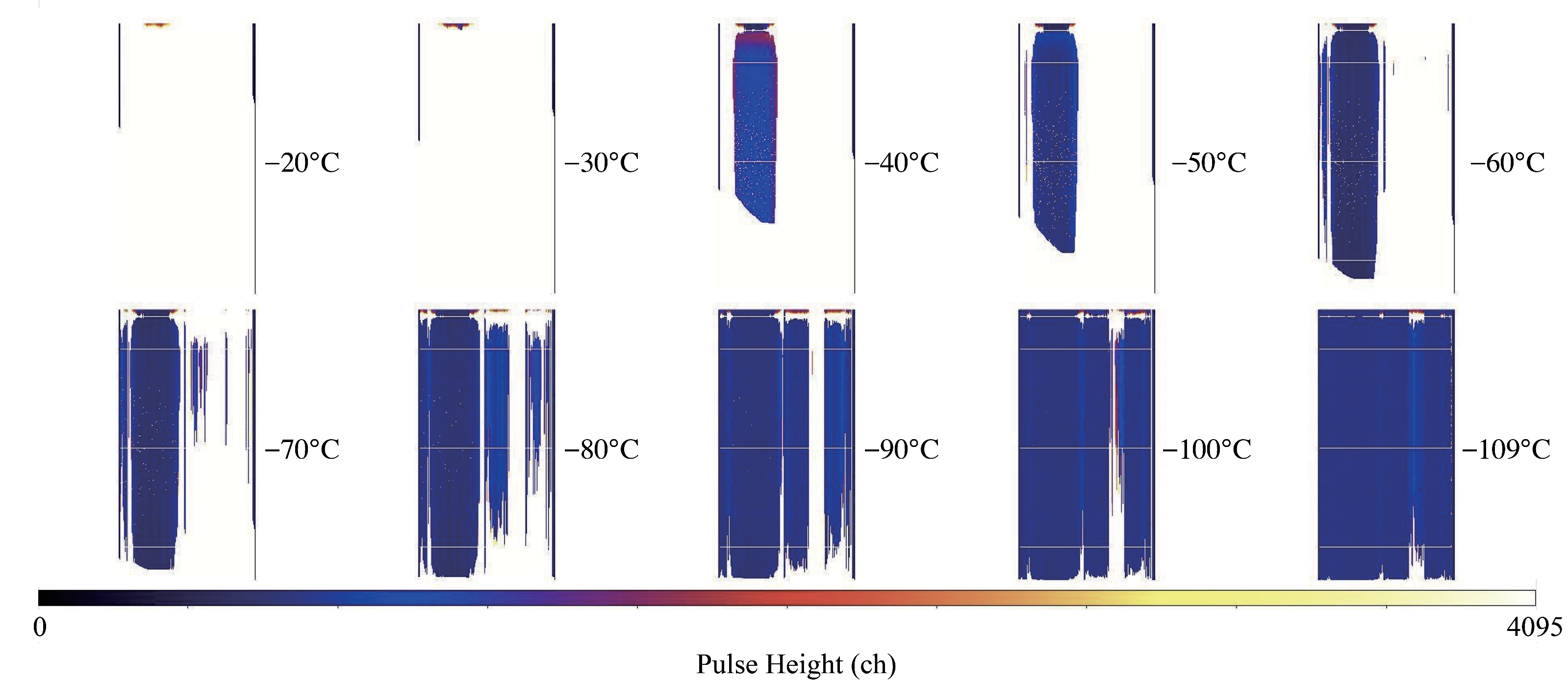}
   \end{tabular}
   \end{center}
   \caption[example] 
   { \label{fig:temp} 
Frame images with the anomalous charge issue captured at various CCD temperatures ranging from $-20~^\circ\mathrm{C}$ to $-109~^\circ\mathrm{C}$. The CCD segment and readout system used are identical to those in Figure~\ref{fig:issue} (right), but these images were obtained from a separate experiment conducted during the SXI cooling test. }
   \end{figure} 

To investigate the cause of anomalous charges, we first focus the temperature dependence of the anomalous charge issue. 
Even under normal conditions without the issue, the pulse heights vary depending on the CCD temperature. When the CCD temperature is above $-30~^\circ\mathrm{C}$, the pulse heights across the entire frame are pinned at 4095~ch due to dark current. 
As the CCD temperature decreases from $-40^\circ\mathrm{C}$ to $-60~^\circ\mathrm{C}$, the pulse heights gradually drop to the pedestal level of $300 - 700$~ch across most of the frame, excluding the edge regions, the CI lines, and some hot or warm pixels. 
When the temperature drops below $-100~^\circ\mathrm{C}$, the entire frame, excluding the CI lines, stabilizes at $300 - 700$~ch.
Therefore,  by considering the pulse height changes under normal conditions, we can identify the anomalous charge issue and its temperature dependence below apploximately $-40~^\circ\mathrm{C}$.

To examine the temperature dependence of the anomalous charge issue, we conducted an experiment in which we varied the CCD temperature during the SXI cooling test in 2020/2021 (\S \ref{sec3}). 
This experiment was performed separately from the one that produced Figure~\ref{fig:issue}~(right), but it used the same CCD segment and readout system. 
Figure~\ref{fig:temp} presents frame images showing the anomalous charge issue at different CCD temperatures.
When the CCD temperature is above $-30~^\circ\mathrm{C}$, the pulse heights across the entire frame are pinned at 4095~ch, making it impossible to distinguish whether the observed effect is due to thermal noise or anomalous charges. 
Below $-40~^\circ\mathrm{C}$, a significant portion of the frame image remains pinned at 4095~ch, clearly due to the anomalous charge issue. As the temperature decreases from $-40~^\circ\mathrm{C}$ to $-90~^\circ\mathrm{C}$, the severity of the anomalous charge issue gradually diminishes. 
During this temperature range, the anomalous charge remains prominent, with the \textit{Mountain} and \textit{Icicle} features still connected in many columns. 
When the temperature drops below $-100~^\circ\mathrm{C}$, the \textit{Mountain} and \textit{Icicle} are separated, making \textit{Field} more prominent, and the frame images resemble that shown in Figure~\ref{fig:issue}(right).
These results demonstrate temperature dependence of the anomalous charge issue, with the amount of anomalous charges decreasing as the CCD temperature decreases.

\subsection{Dependence on Potential Structures Created by CI Electrodes}
\label{sec5p2}

We next focus on the dependence of the anomalous charge issue on the structure of the potential barrier.
The new CCD driving technique (\S\ref{sec3}) sets the potential of P1V higher than those of IG1V and IG2V to create a potential barrier, successfully preventing anomalous charge intrusion from outside IA, as shown in \S\ref{sec4}. However, it remains unclear whether anomalous charges are generated around the potentials of IG1V and IG2V (case A) or further outside (case B) because the potential barrier can be effective in both scenarios. To distinguish between cases A and B, the key lies in the dependence of the amount of anomalous charges appearing in frame images on the potentials of IG1V and IG2V. Therefore, we conducted a cooling test to examine how the degree of the anomalous charge issue varies with changes in the potential structures of IG1V and IG2V. After reproducing the anomalous charge issue in the non-FM system (\S\ref{sec4}), the potentials of IG1V and IG2V were individually changed from minimum to maximum, and frame images were inspected. Note that the new CCD driving technique was not applied, but the normal operation was conducted in this experiment. 
Furthermore, we kept the potential of ISV constant to make the CI technique not applied because we focused on not the spectroscopic performance but on changes in the anomalous charge issue.

   \begin{figure} [t]
   \begin{center}
   \begin{tabular}{c} 
   \includegraphics[height=12.5cm]{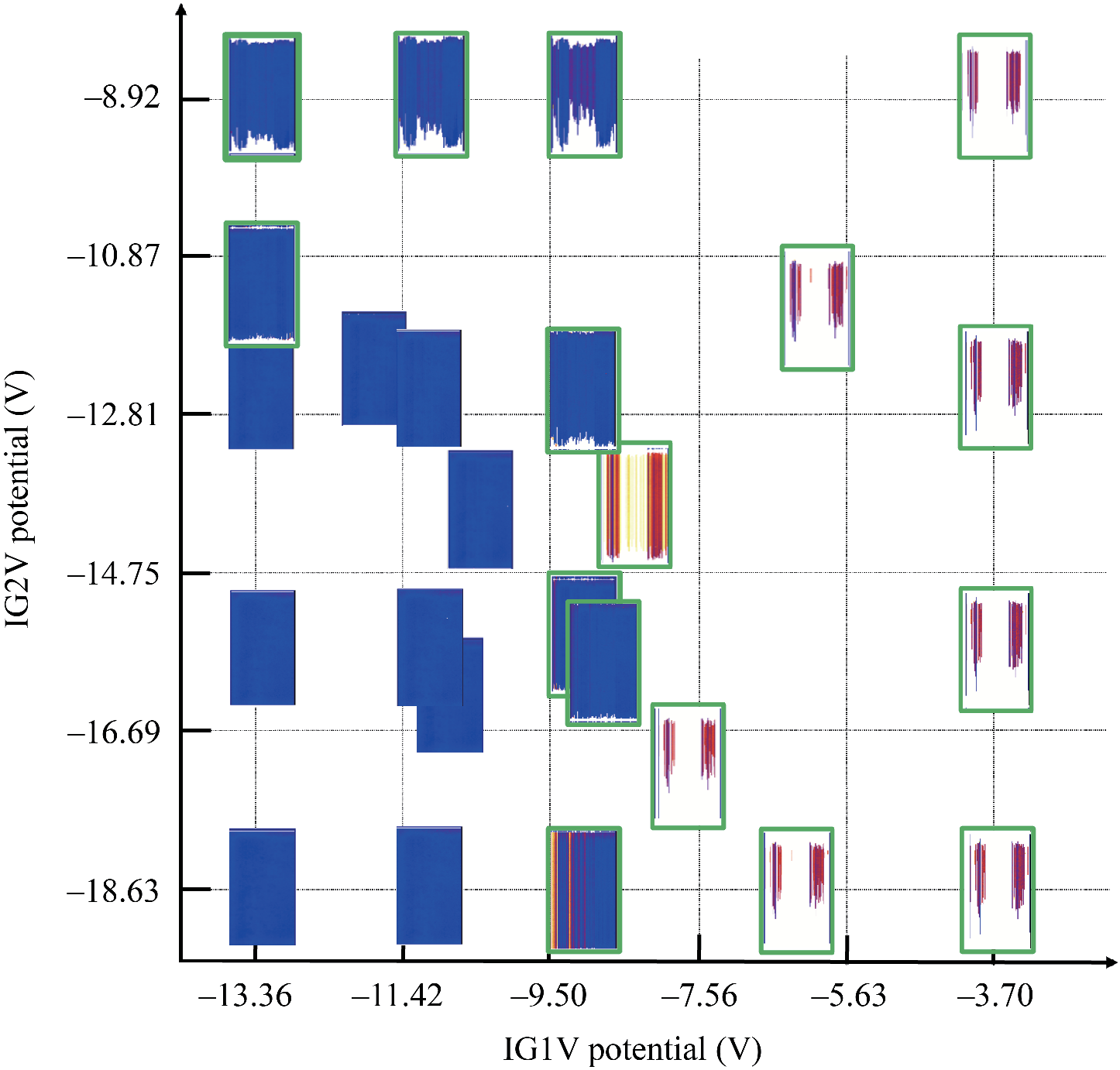}
   \end{tabular}
   \end{center}
   \caption[example] 
   { \label{fig:cause} 
The IG1V$-$IG2V plane and the corresponding frame images when the CCD temperature is approximately $-45~^{\circ}$C. The frame images, where the anomalous charge issue is observed, are enclosed by green squares.}
   \end{figure} 

Figure \ref{fig:cause} shows how frame images change on the IG1V$-$IG2V plane. When the potential of IG1V is held constant at the minimum and the potential of IG2V is varied from the minimum to maximum, the amount of anomalous charges begins to increase around IG1V $\sim -10.87$~V. Conversely, if the IG1V potential is varied from the minimum to maximum with the IG2V potential fixed at the minimum, anomalous charges start to increase at IG1V $\sim -9.50$~V. Although the amount of anomalous charges tends to increase regardless of whether IG1V or IG2V is increased, the degree of the anomalous charge issue is more prominent when the potential of IG1V is at its maximum compared to when that of IG2V is at its maximum.
Field emission is unlikely to explain these results, as it has been confirmed that when the anomalous charge issue does not occur (Figure~\ref{fig:issue} left), varying the potentials of IG1V or IG2V does not affect pulse heights in the IA.
Within the ranges of $-18.63 \lesssim \textrm{IG1V} \lesssim -12.81$ and $-13.36 \lesssim \textrm{IG2V} \lesssim -11.42$, no anomalous charges appeared in the frame images, and the pedestal levels of pulse heights remained within the range of $300 - 700$~ch.

If the source of the anomalous charges is located outside the potential of IG1V, the maximum potential of IG1V should create the highest potential barrier against them. However, the results in Figure~\ref{fig:cause} seem to contradict this idea, as a higher potential of IG1V allows a larger amount of anomalous charges to intrude into IA. This suggests that the source of the anomalous charges might be located near the potentials of IG1V or IG2V (Case A). Since the potential of IG2V has less impact on the degree of the anomalous charge issue compared to IG1V, the region around the potential of IG1V appears more likely.
To conclusively determine the location of the anomalous charge source and identify the underlying physical mechanism, further studies are needed to explore the dependence on other parameters, such as the bias level, reset and/or output drain voltages, and the isolation of the IA from the edges of the chip. These will be addressed in future work. 

\section{Summary}
\label{sec6}

In the development of the X-ray CCD camera, SXI, for Xtend onboard XRISM, we devised a new CCD driving technique to prevent the anomalous charge issue that occurred during the SXI cooling test in 2020/2021, in case it were to happen in orbit in the future.
We reproduced the issue by combining the CCD chips that experienced it with the non-FM system and confirmed that anomalous charges can be successfully prevented by the new technique. 
Furthermore, during the satellite thermal vacuum test in 2022, we also confirmed that the imaging and spectroscopic performance with the new technique is comparable to that with normal operation. 
To identify the cause of the anomalous charges, we examined frame images by varying the CCD temperature and the potential structures by IG1V and IG2V. As a result, we found that the severity of the issue decreased with lower CCD temperatures and increased with higher potentials of IG1V and IG2V. Further studies on the dependence of various parameters, in addition to the CCD temperature and the potential structures created by the CI electrodes, are needed to coclude the exact location of the source of anomalous charges and to understand the physical mechanism behind their generation.

\section*{DATA AVAILABILITY}
The data that support the findings of this article are proprietary and are not publicly available. 
The data plotted in the above figures are available from the corresponding author upon request, and a limited subset of the underlying data  can be requested from the author at hirofumi.noda@astr.tohoku.ac.jp.

\section*{DISCLOSURES}

The authors declare there are no financial interests, commercial affiliations, or other potential conflicts of interest that have influenced the objectivity of this research or the writing of this paper.

\acknowledgments 
 
This work is supported by Japan Society for the Promotion of Science (JSPS) KAKENHI with the Grant number of 19K21884, 20H01941, 20H01947, 20KK0071, 21H01095, 21J00031,  21K20372, 22KJ3059, 23K20239, 23K20850, 23K22536, 24K00672.


\end{document}